\documentstyle[12pt,a4,epsfig]{article}
\newcommand{\bm}[1]{\mbox{\boldmath $#1$}}
\newcommand{\bt}[1]{{\bm #1}_{_T}}
\newcommand{\Pslash}{\kern 0.2 em P\kern -0.56em \raisebox{0.3ex}{/}}
\newcommand{\pslash}{\kern 0.2 em p\kern -0.4em /}
\newcommand{\nslash}{\kern 0.2 em n\kern -0.56em /}
\newcommand{\kslash}{\kern 0.2 em k\kern -0.45em /}
\newcommand{\Sslash}{\kern 0.2 em S\kern -0.56em \raisebox{0.3ex}{/}}
\newcommand{\dslash}{\kern 0.2 em \partial\kern -0.56em \raisebox{0.3ex}{/}}
\newcommand{\xbj}{x_{_B}}

\newcommand{\bkt}{\bt{k}}
\newcommand{\bpt}{\bt{p}}
\newcommand{\bqt}{\bt{q}}
\newcommand{\bSt}{\bt{S}}
\newcommand{\be}{\begin{equation}}
\newcommand{\ee}{\end{equation}}
\newcommand{\bea}{\begin{eqnarray}}
\newcommand{\eea}{\end{eqnarray}}
\newcommand{\st}{{\scriptscriptstyle T}}
\begin{document}

\begin{flushright}
hep-ph/9710498\\
NIKHEF 97-044\\
VUTH 97-19
\end{flushright}
\begin{center}
{\LARGE\bf Intrinsic transverse momentum \\ 
\quad \\
in hard processes}\footnote{
Talk presented at the Symposium "Deep inelastic scattering
off polarized targets: theory meets experiment", Sep. 1-5, 1997, Zeuthen}

\vspace{1cm}
{P.J. Mulders}

\vspace*{1cm}
{\it Department of Physics and Astronomy, Free University} \\
and \\
{\it National Institute for Nuclear Physics and High Energy Physics (NIKHEF)\\
P.O. Box 41882, NL-1009 DB Amsterdam, The Netherlands}\\

\vspace*{2cm}

\end{center}

\begin{abstract}
In this contribution I discuss transverse momentum dependent quark
distributions and their operator structure. Some of the experimental 
signatures in particular in Drell-Yan scattering are given.
\end{abstract}

\section{Introduction}

\vspace{1mm}
\noindent
In hard processes one becomes sensitive to intrinsic transverse momentum
of the quarks if at least three external momenta are present.
In Drell-Yan (DY) scattering ($A B \rightarrow \mu^+\mu^-X$) one has the 
momenta 
of the incoming hadrons $A$ and $B$ and the hard time-like momentum of 
the lepton-pair $q$ (with $q^2$ = $Q^2$). Other examples are 1-particle
inclusive leptoproduction ($\ell H \rightarrow \ell^\prime h X$) and 
two-hadron production in $e^+e^-$ annihilation.

In a field-theoretic approach one can calculate the hadronic tensor
$W_{\mu\nu}$ for DY scattering by making use of a diagrammatic expansion
in which soft parts allow inclusion of hadrons in the field-theoretical
framework of QCD. If $Q^2 \rightarrow \infty$ only a limited number of
diagrams survives, each of these diagrams with characteristic behavior. 
Some explicit
examples are shown below. In particular the tree level diagram in the first
box and the diagrams with a gluon attached to the soft part 
(one shown in the second box) will be discussed in more detail.
\begin{center}
\fbox{
\begin{minipage}{7.0 cm}
\epsfxsize=5.5cm \epsfbox{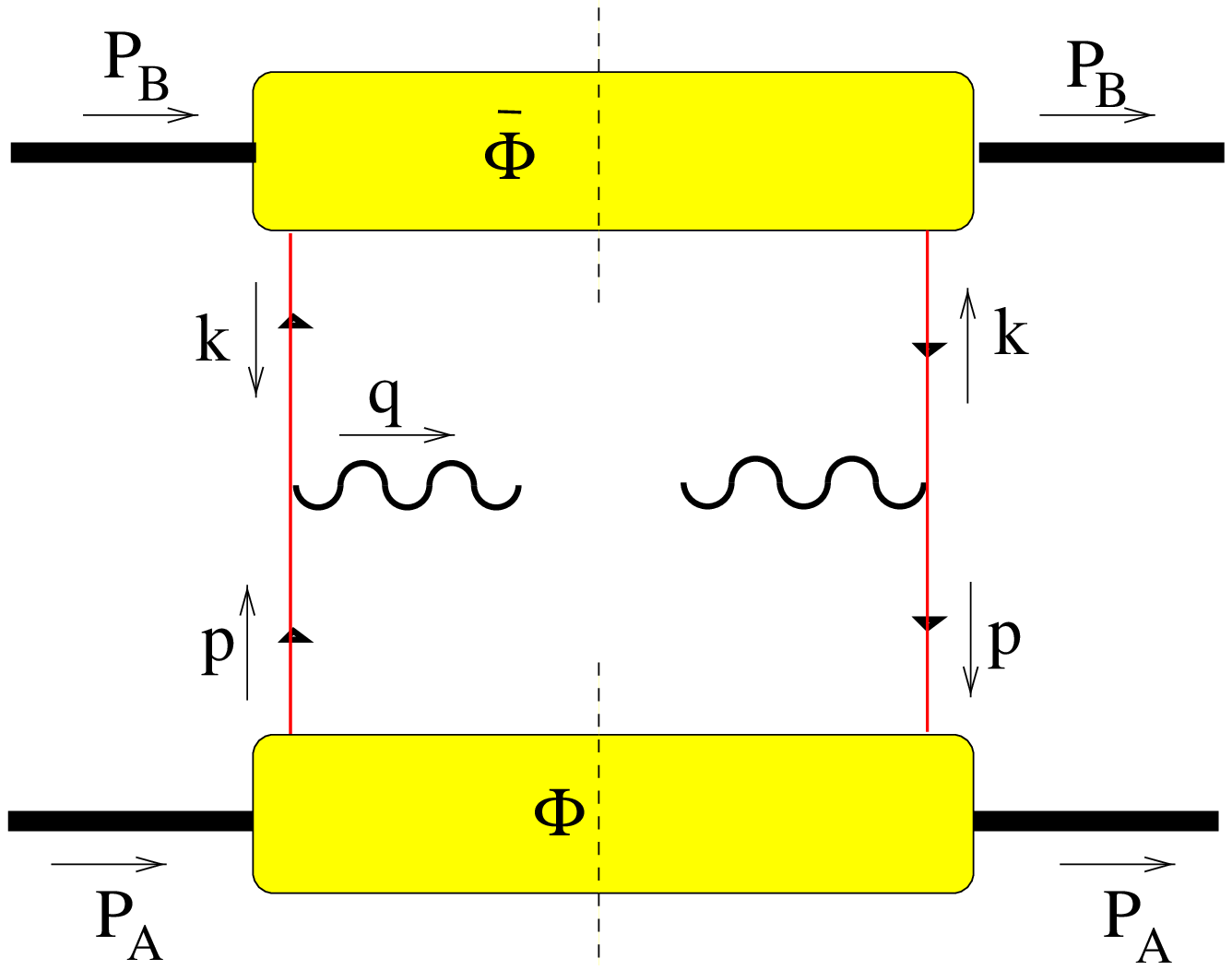}
\end{minipage}
\begin{minipage}{8.0 cm}
Parton model result \\
(scales)
\end{minipage}
}
\end{center}
\begin{center}
\fbox{
\begin{minipage}{7.0 cm}
\epsfxsize=5.5cm \epsfbox{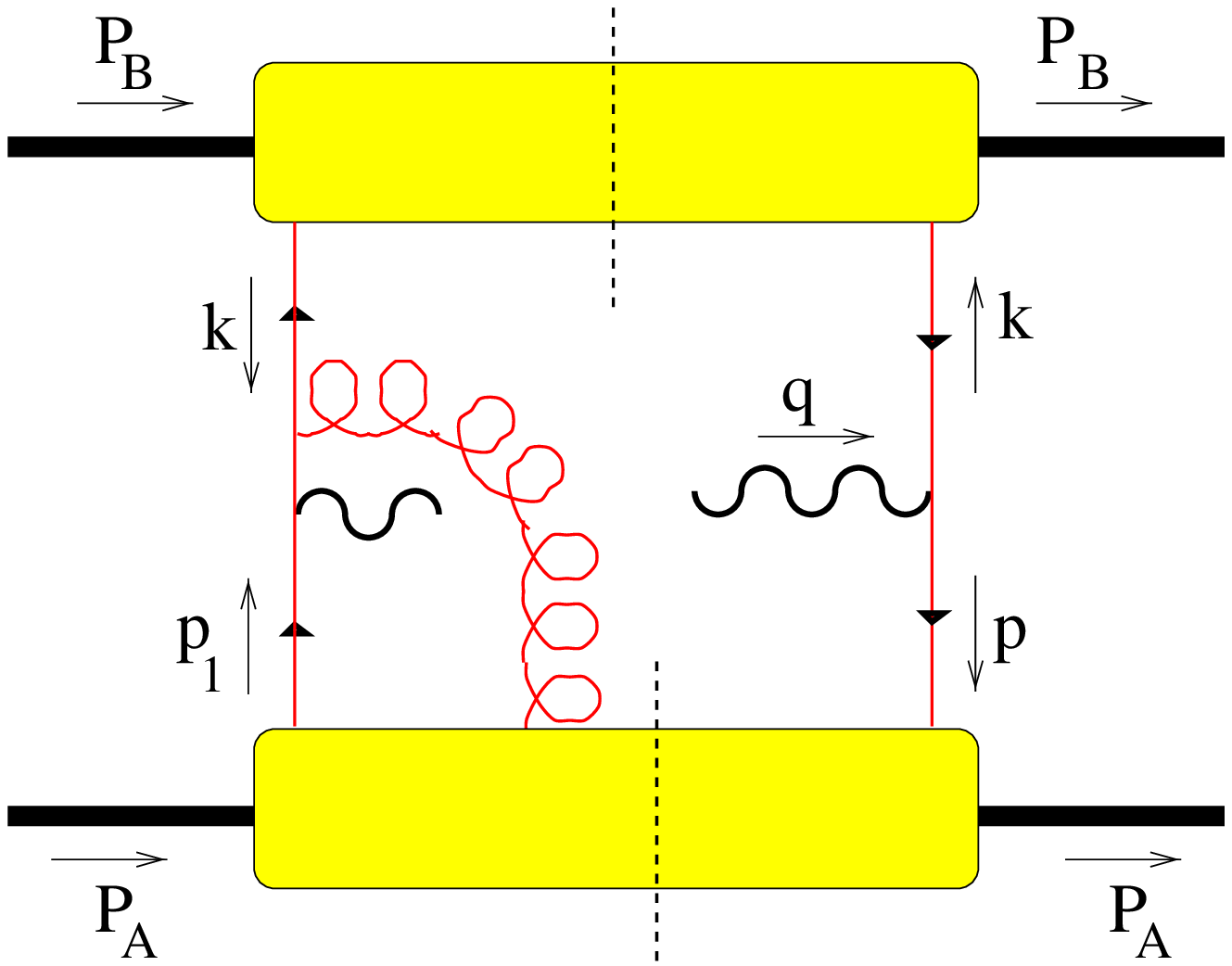}
\end{minipage}
\begin{minipage}{8.0 cm}
\begin{itemize}
\item
'Physical' $A_T$ gluons\\
{}[e.m. gauge invariance at ${\cal O}(1/Q)$]
\item
Longitudinal $A^+$ gluons\\
{}[produce link, color gauge invariance]
\end{itemize}
\end{minipage}
}
\end{center}
\begin{center}
\fbox{
\begin{minipage}{11.5 cm}
\epsfxsize=5.5cm \epsfbox{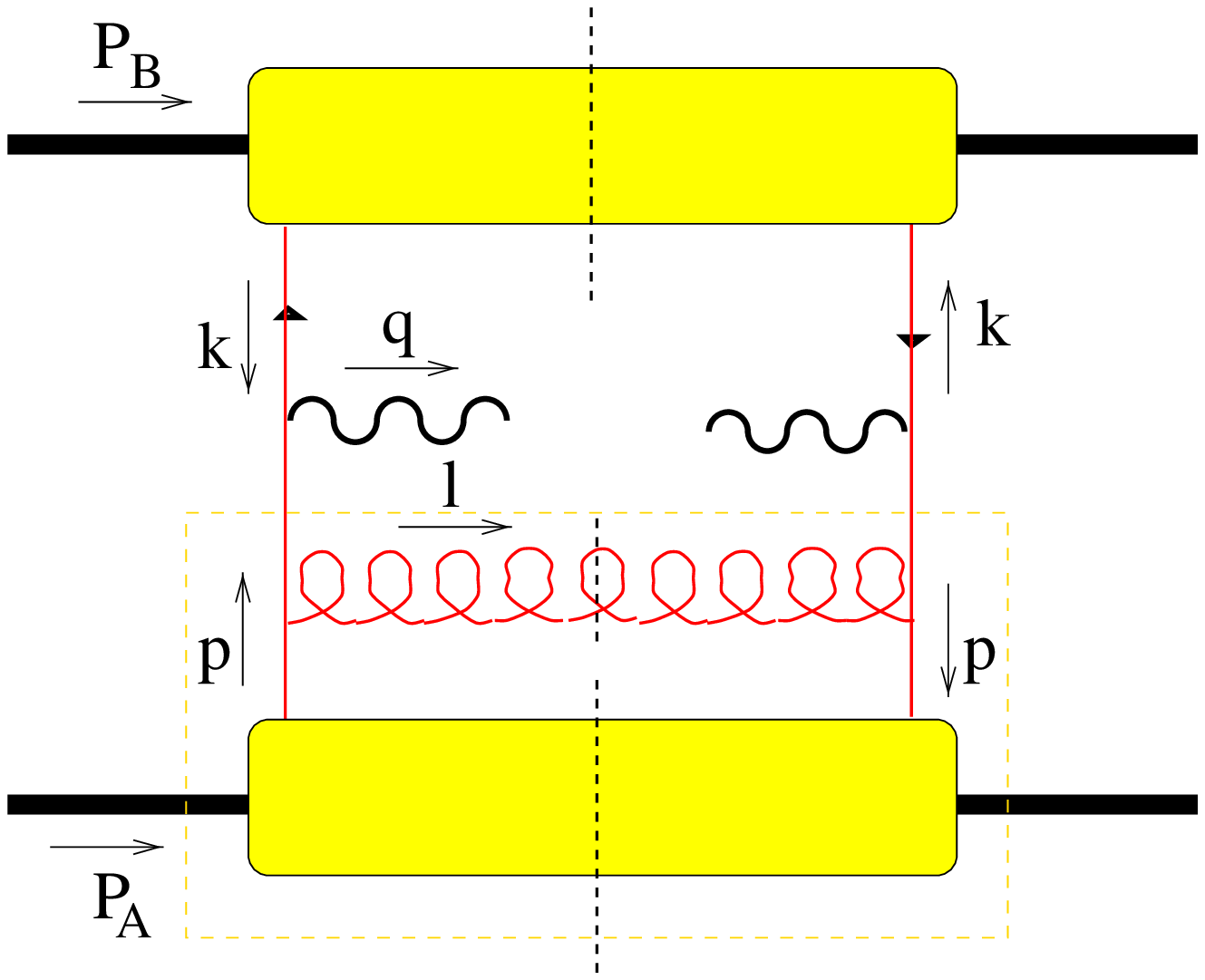}
\epsfxsize=5.5cm \epsfbox{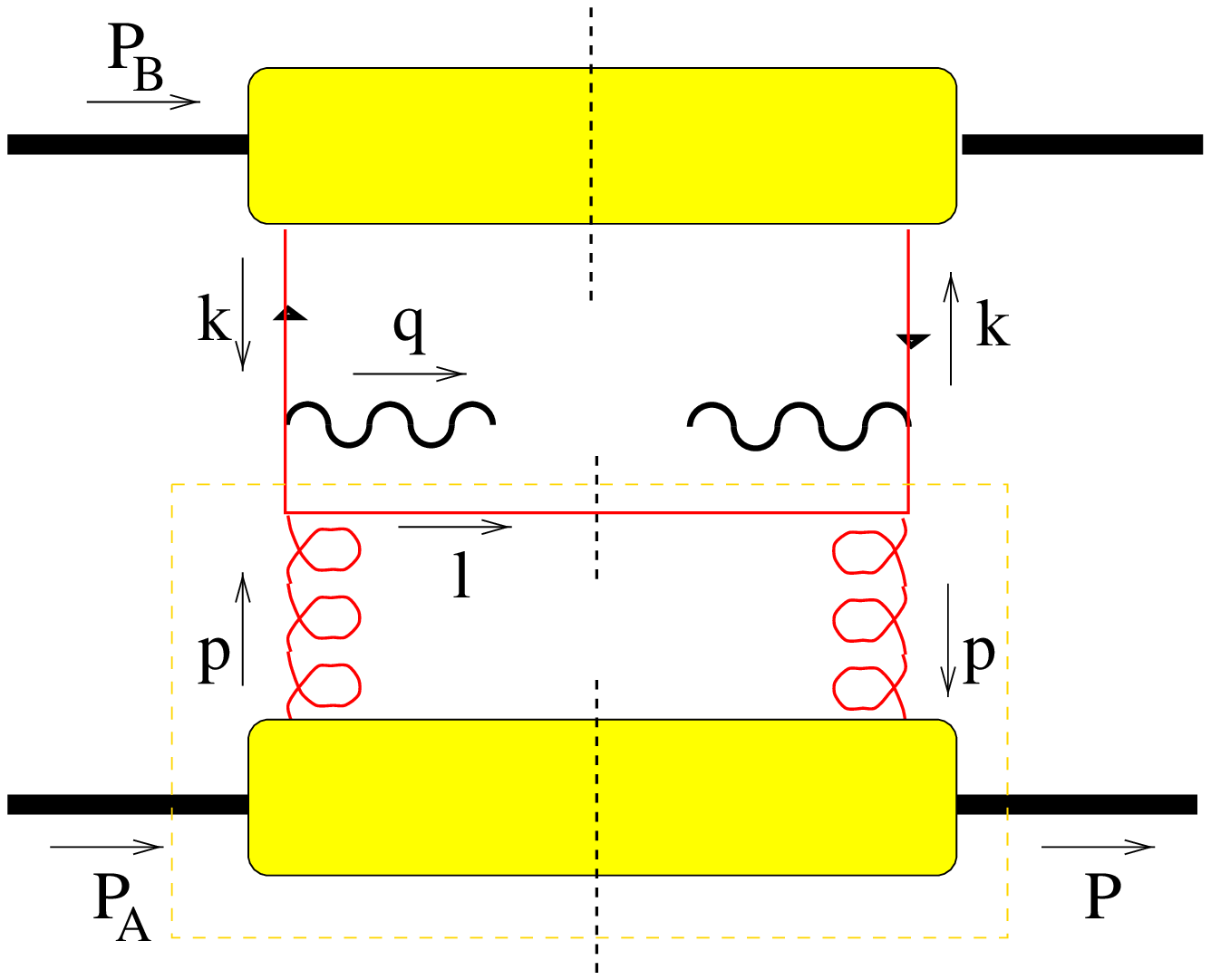}
\end{minipage}
\begin{minipage}{3.5 cm}
$\alpha_s \log(Q^2/\mu^2)$\\
corrections\\
{}[DGLAP equations]
\end{minipage}
}
\end{center}
\begin{center}
\fbox{
\begin{minipage}{7.0 cm}
\epsfxsize=5.5cm \epsfbox{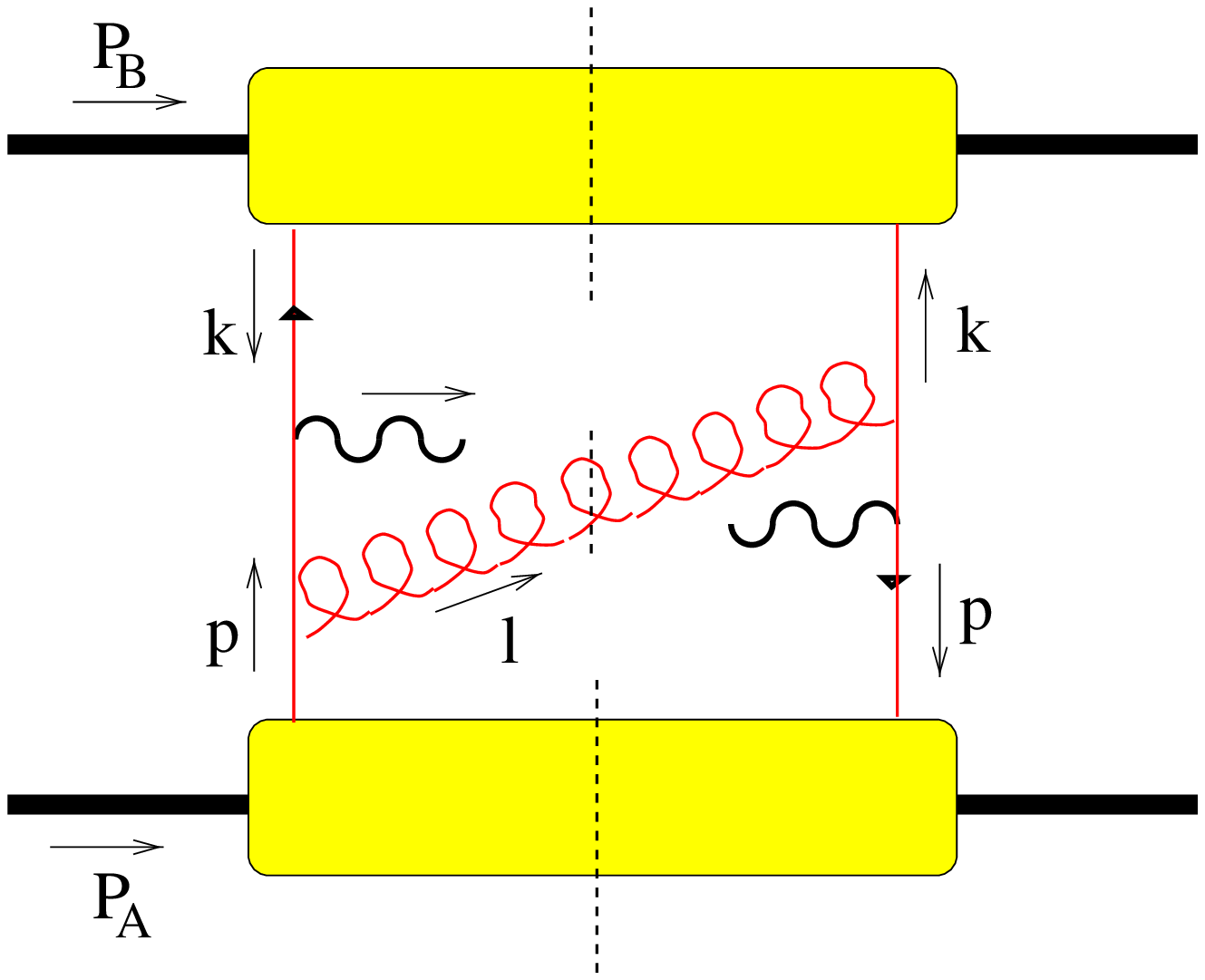}
\end{minipage}
\begin{minipage}{8.0 cm}
(Process dependent) $\alpha_s$ corrections
\end{minipage}
} 
\end{center}

\section{Soft parts}

\vspace{1mm}
\noindent
The soft parts that appear in the diagrammatic expansion are the quark-quark
correlation function, 
\vspace{0.2 cm}
\\
\begin{minipage}{7.0 cm}
\epsfxsize=5.5cm \epsfbox{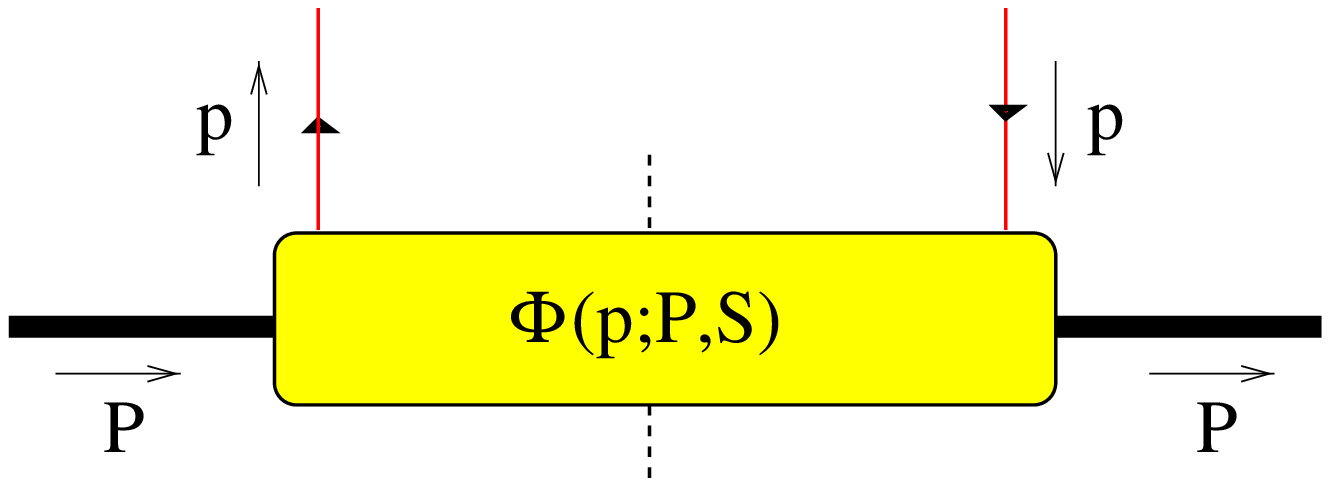}
\end{minipage}
\begin{minipage}{7 cm}
\[
p^2 \sim p\cdot P \sim P^2 = M^2 \ll Q^2
\]
\end{minipage}
\\
given by
\be
\Phi_{ij}(p;P,S)=\frac{1}{(2\pi)^4}\int d^4x \;
e^{ip\cdot x}\;\langle P,S|{\overline\psi_j(0)\psi_i(x)}|P,S\rangle ,
\ee
and the quark-quark-gluon correlation function
\vspace{0.2 cm}
\\
\begin{minipage}{6.0 cm}
\epsfxsize=5.5cm \epsfbox{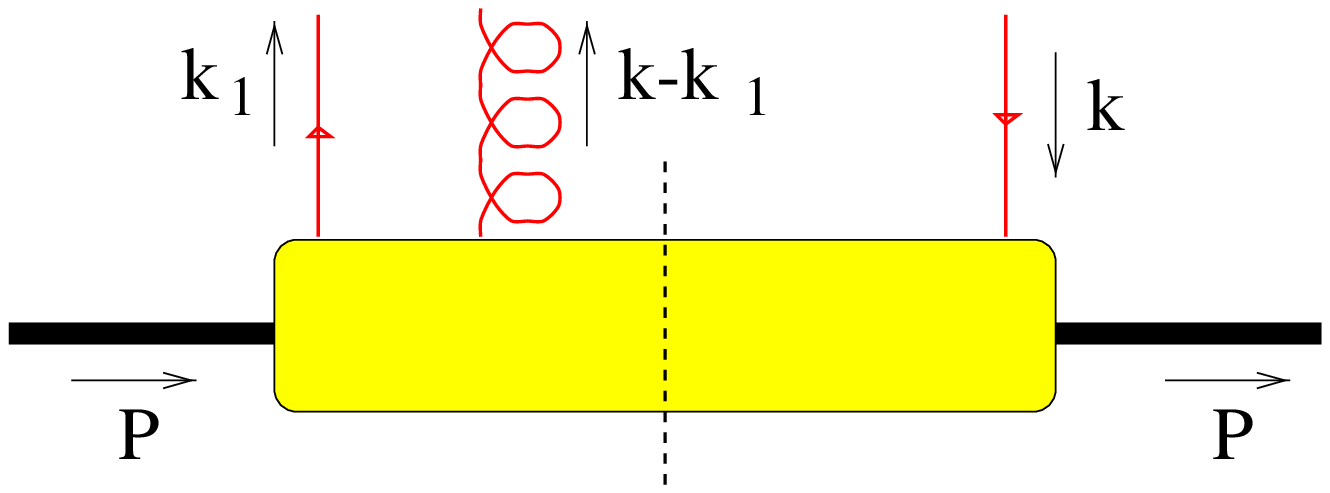}
\end{minipage}
\begin{minipage}{8.0 cm}
\begin{eqnarray*}
&&k^2 \sim k\cdot P \sim k_1^2 \sim k_1\cdot P  \sim k\cdot k_1
\\ &&\hspace{2.5 cm} \mbox{}
\sim  P^2 = M^2 \ll Q^2
\end{eqnarray*}
\end{minipage}
\\
given by
\be
\Phi^\alpha_{A\,ij}(k,k_1;P,S)  =  \frac{1}{(2\pi)^4}\int d^4x\,d^4y\
e^{i\,k_1\cdot
(x-y) + ik\cdot y} 
\ \langle P,S \vert 
{\overline \psi_j(0)\,gA_T^\alpha (y) \psi_i(x)} \vert P,S \rangle .
\label{gluonblob}
\ee

The large scale $Q$ in the scattering process leads in a natural way 
to the use of lightlike vectors for which
$n_+^2 = n_-^2$ = 0 and $n_+\cdot n_-$ = 1.
For the DY process:
\[
\left.
\begin{array}{l} q^2 = Q^2 \\
P_A^2 = M_A^2\\
P_B^2 = M_B^2 \\
2\,P_A\cdot q = Q^2/x_{_A} \\
2\,P_B\cdot q = Q^2/x_{_B} \\
\end{array} \right\}
\longleftrightarrow \left\{
\begin{array}{l}
P_B = \frac{Q}{x_{_B}\sqrt{2}}\,n_-
+ {\frac{x_{_B} M_B^2}{Q\sqrt{2}}}\,n_+
\\ \mbox{} \\
q =\frac{Q}{\sqrt{2}}\,n_- + \frac{Q}{\sqrt{2}}\,n_+ + {q_\st}
\\ \mbox{} \\
P_A = {\frac{x_{_A} M_A^2}{Q\sqrt{2}}}\,n_-
+ \frac{Q}{x_{_A} \sqrt{2}}\,n_+
\end{array}
\right.
\]
Using the lightlike vectors to define lightcone components $p^\pm 
= p\cdot n_\mp$ one sees that the relevant quantities entering in the
hard process for the parton graph in the first box above
are $\int dp^- \,\Phi$ and $\int dk^+ \,\overline \Phi$. The structure
of these quantities, their various Dirac projections, and a possible
way of estimating their magnitude has been
discussed in the contribution of Jakob~\cite{JMR97}.
The relevant part in $\int dp^-\,\Phi$,
\bea
\Phi_{ij}(x,\bm p_\st) & = &
\int {dp^-} \ \Phi_{ij}(p;P,S) 
\nonumber
\\ & = &
\left. \int \frac{d\xi^-d^2\bm \xi_\st}{(2\pi)^3}\ e^{ip\cdot \xi}
\,\langle P,S\vert {\overline \psi_j(0)} {\psi_i(\xi)}
\vert P,S\rangle \right|_{\xi^+ = 0},
\eea
which is a light-front correlation function, in leading order in $1/Q$ 
given by~\cite{RS80,TM95}
\bea
\Phi(x,\bm p_\st) & = &
\frac{1}{2}\Biggl\{{f_1}\nslash_+
+ {\lambda}\,{g_{1L}}\gamma_5\nslash_+
+ {g_{1T}}\,\frac{(\bpt\cdot\bSt)}{M}\,\gamma_5\nslash_+
+ {h_{1T}}\,\frac{[{\Sslash_\st},\nslash_+]\gamma_5}{2}
\nonumber \\ && \quad \mbox{}
+ {\lambda}\,{h_{1L}^\perp}
\,\frac{[\pslash_\st,\nslash_+]\gamma_5}{2}
+ {h_{1T}^\perp}\,\frac{(\bpt\cdot\bSt)}{M}
\,\frac{[\pslash_\st,\nslash_+]\gamma_5}{2}\Biggr\}
+ {\cal O}\left(\frac{M}{P^+}\right),
\eea
where $p^+ = x\,P^+$ and the functions $f_1 = f_1(x,\bpt^2)$ etc. The
polarization vector is also expanded with the help of the lightlike
vectors,
\be
S = -{\lambda}\frac{M}{2P^+}\,n_-
+ {\lambda}\frac{P^+}{M}\,n_+ + {S_\st}.
\ee
and satisfies $\lambda^2 + \bm S_\st^2 = 1$ (for a pure state).
The functions appearing in $\Phi(x,\bpt)$ can be projected out with
specific Dirac matrices. All functions in the leading part can be
interpreted as densities involving chirally left/right or transverse
spin projections of good quark fields, 
$\psi_+ \equiv (\gamma^-\gamma^+/2)\psi$.

In inclusive lepton-hadron scattering or $\bqt$-integrated DY
scattering one does not need the $\bpt$-dependent quark distributions 
discussed above, but just the $\bpt$-averaged results, i.e. 
\bea
\Phi_{ij}(x) & = &
\int {dp^-\,d^2\bm p_\st} \ \Phi_{ij}(p;P,S)  
\nonumber \\ & = &
\left. \int \frac{d\xi^-}{2\pi}\ e^{ip\cdot \xi}
\,\langle P,S\vert {\overline \psi_j(0)} {\psi_i(\xi)}
\vert P,S\rangle \right|_{LC},
\eea
where 'LC' denotes $\xi^+ = \xi_\st = 0$, which implies
$\xi^2$ = 0, i.e.\ one deals with a lightcone correlation function.
In leading order one has
\be
\Phi(x) =
\frac{1}{2}\Biggl\{{f_1}(x)\nslash_+
+ {\lambda}\,{g_{1}}(x)\,\gamma_5\nslash_+
+ {h_{1}}(x)\,\frac{[{\Sslash_\st},\nslash_+]\gamma_5}{2}
\Biggr\} + {\cal O}\left(\frac{M}{P^+}\right),
\ee
where ${f_1(x)} = \int d^2\bm p_\st\ {f_1(x,\bm p_\st)}$, etc.
For asymmetries in 1-particle inclusive lepton-hadron scattering or in
DY scattering one also needs the 
$\bpt$-weighted results,
\bea
\Phi_{\partial\,ij}^{\alpha}(x) & = &
\int {dp^-\,d^2\bm p_\st \,\,p_\st^\alpha}\,\Phi_{ij}(p;P,S) 
\nonumber \\ & = &
\left. \int \frac{d\xi^-}{2\pi}\ e^{ip\cdot \xi}
\,\langle P,S\vert {\overline \psi_j(0)} 
{i\partial^\alpha\psi_i(\xi)} \vert P,S\rangle \right|_{LC},
\eea
for which one in leading order has
\be
\frac{1}{M}\,\Phi_\partial^{\alpha}(x) =
\frac{1}{2}\,\Biggl\{
{g_{1T}^{(1)}}(x)\,{S_\st^\alpha}\,\gamma_5\nslash_+
+{\lambda}\,{h_{1L}^{\perp (1)}}(x)
\,\frac{[\nslash_+, \gamma^\alpha]\gamma_5}{2}\Biggr\}
+ {\cal O}\left(\frac{M}{P^+}\right),
\ee
with e.g.\ ${g_{1T}^{(1)}(x)}$
$\equiv$ $\int d^2\bm p_\st\,(\bpt^2/2M^2)\,{g_{1T}(x,\bpt)}$.

Including subleading order for the $\bpt$-averaged functions
we just quote the result
\bea
\Phi(x) & = &
\frac{1}{2}\Biggl\{{f_1}(x)\,\nslash_+
+ {\lambda}\,{g_{1}}(x)\,\gamma_5\nslash_+
+ {h_{1}}(x)\,\frac{[{\Sslash_\st},\nslash_+]\gamma_5}{2}
\Biggr\}
\\ && \mbox{}+\frac{M}{2P^+}\Biggl\{
{e}(x) + {g_T}(x)\,\gamma_5{\Sslash_\st} 
+ {\lambda}\,{h_L}(x)\,\frac{[\nslash_+,\nslash_-]\gamma_5}{2} 
\Biggr\}
+ {\cal O}\left(\frac{M}{P^+}\right)^2.
\eea
The factor $M/P^+$ leads in a calculation to a suppression factor
$M/Q$, hence the functions $e$, $g_T$ and $h_L$ appear at
subleading order.
These functions do not have a simple density interpretation.
From the most general amplitude expansion for $\Phi_{ij}$, however,
constrained by hermiticity, parity and time reversal symmetry one 
obtains relations between the twist-three functions and the 
$\bpt$-dependent distribution functions~\cite{BKL84,MT95},
\bea
&&
\underbrace{g_T(x) - g_1(x)}_{g_2(x)} \ =\  \frac{d}{dx}\,g_{1T}^{(1)},
\label{rel1}
\\ &&
\underbrace{h_L(x) - h_1(x)}_{{1\over 2}\,h_2(x)} 
\ =\  -\,\frac{d}{dx}\,h_{1L}^{\perp(1)}.
\label{rel2}
\eea

\section{Gluons in the diagrammatic calculation}

\vspace{1mm}
\noindent
The soft parts discussed in the previous section and the analogous
results for antiquarks are sufficient to calculate the parton diagram
(first box in the introduction). Turning to the second box one finds
diagrams with a gluon attached to the soft part involving the matrix elements
of Eq.~\ref{gluonblob}. These contributions are important at 
subleading order~\cite{JJ92}.
The nonlocality up to ${\cal O}(1/Q)$ can be
restricted to a lightlike nonlocality involving
$\overline \psi(0)A_T^\alpha(\xi)\,\psi(\xi)$ with $\xi^+ = \xi_\st = 0$.
At least, this is true in the absence of so-called gluonic 
poles~\cite{BMT97}, which is the case when gluonic background matrix
elements having a nonlocality of the type 
$\overline \psi(0)A_T^\alpha(\eta^- = \infty)\,\psi(\xi^-)$ vanish.

To ensure gauge invariance one looks for the combination 
$p_\st^\alpha + g\,A_T^\alpha$, i.e.\ for quark-quark correlation functions
containing the covariant derivative
$iD_T^\alpha = i\partial_\st^\alpha + g\,A_T^\alpha$
\bea
\Phi_{A\,ij}^{\alpha}(x) & = &
\left. \int \frac{d\xi^-}{2\pi}\ e^{ip\cdot \xi}
\,\langle P,S\vert {\overline \psi_j(0)} 
{g\,A_T^\alpha(\xi)\,\psi_i(\xi)}
\vert P,S\rangle \right|_{LC}
\nonumber \\ & = &
\Phi_{D\,ij}^{\alpha}(x)
-\Phi_{\partial\,ij}^{\alpha}(x).
\eea
In the next step one employs
the equations of motion,
$\left(i\gamma^\mu D_\mu - m\right)\psi = 0$,
to relate $\Phi_D$ to $\Phi$. In this way one obtains
\bea
\Phi_{D}^{\alpha}(x) &=&
\frac{M}{2}\,\Biggl\{
\left( x\,{g_T} - \frac{m}{M}\,{h_1}\right)
{S_\st^\alpha}\,\gamma_5\nslash_+
+\lambda \left( x\,{h_L} - \frac{m}{M}\,{g_1}\right)
\frac{[\gamma^\alpha,\nslash_+]\gamma_5}{4}
\nonumber \\ && \qquad \quad \mbox{}
+\left( x\,e - \frac{m}{M}\,{f_1}\right)
\frac{[\nslash_+,\gamma^\alpha]}{2}
\Biggr\}.
\eea
In the diagrammatic calculation up to ${\cal O}(1/Q)$ the parton graph 
and the gluon graphs ensure electromagnetic gauge invariance~\cite{JJ92}.
The difference between $\Phi_D^\alpha$ and $\Phi_\partial^\alpha$ gives
the truly interaction dependent part,
\be
\Phi_{A}^{\alpha}(x)=
\frac{M}{2}\,\Biggl\{
x\,{\tilde g_T(x)}\,{S_\st^\alpha}\,\gamma_5\nslash_+
+\lambda\,x\,{\tilde h_L(x)}
\,\frac{[\gamma^\alpha,\nslash_+]\gamma_5}{4}
+x\,\tilde e(x) \,\frac{[\nslash_+,\gamma^\alpha]}{2}
\Biggr\},
\ee
with
\bea
&&
x\,{\tilde g_T} =
x\,{g_T} - {g_{1T}^{(1)}} - \frac{m}{M}\,{h_1},
\\ &&
x\,{\tilde h_L} =
x\,{h_L} - {h_{1L}^{\perp(1)}} - \frac{m}{M}\,{g_1},
\\ &&
x\,{\tilde e} =
x\,e - \frac{m}{M}\,{f_1}.
\eea
As an aside we note that one obtains the Wandzura-Wilczek 
relation~\cite{WW77} for $g_T$ and a similar relation~\cite{JJ92}
for $h_L$,
\bea
&&g_T(x) = \int_x^1 dy\,\frac{g_1(y)}{y}
+ \frac{m}{M}\,
\left[ \frac{h_1(x)}{x} - \int_x^1 dy\,\frac{h_1(y)}{y^2}\right],
\\ &&
h_L(x) = 2x\int_x^1 dy\,\frac{h_1(y)}{y^2} 
+ \frac{m}{M} \left[ \frac{g_1(x)}{x}
- 2x \int_x^1 dy\,\frac{g_1(y)}{y^3} \right],
\eea
by using the relations in Eqs~\ref{rel1} and \ref{rel2} and assuming 
$\Phi_A^\alpha$ = 0. 

Included in the diagrammatic expansion are also quark-quark-gluon
correlations involving $A^+$ gluons. These
longitudinal $A^+$ gluons cause problems in the $M/P^+$ expansion.
Their contributions, however, can be collected into the definition of
the distribution functions. One obtains for instance
\be
{f_1}(x) =
\left. \int \frac{d\xi^-}{4\pi}\ e^{ip\cdot \xi}
\,\langle P,S\vert {\overline \psi(0)} {\gamma^+}
\,{\cal P} e^{-ig\int d\eta^\mu A_\mu(\eta)}{\psi(\xi)}
\vert P,S\rangle \right|_{LC},
\ee
rendering the definition of $f_1$ color gauge invariant. In cases
such as inclusive deep inelastic lepton-hadron scattering all
complications can simply be avoided by working in the lightcone
gauge ($A^+$ = 0).

Including longitudinal gluons and transverse momenta we get
\be 
\Phi_{ij}(x,\bm p_\st) =
\int \frac{d\xi^-d^2\bm \xi_\st}{2\,(2\pi)^3}\ e^{ip\cdot \xi}
\,\langle P,S\vert {\overline \psi_j(0)}{{\cal U}(0,\infty)}
\,{\cal U}(\infty,\xi^-)
{\psi_i(\xi)} \vert P,S\rangle \bigr|_{\xi^+ = 0},
\ee
where link operators at fixed transverse positions, 
\begin{center}
\begin{minipage}{7 cm}
\epsfxsize=6.0cm \epsfbox{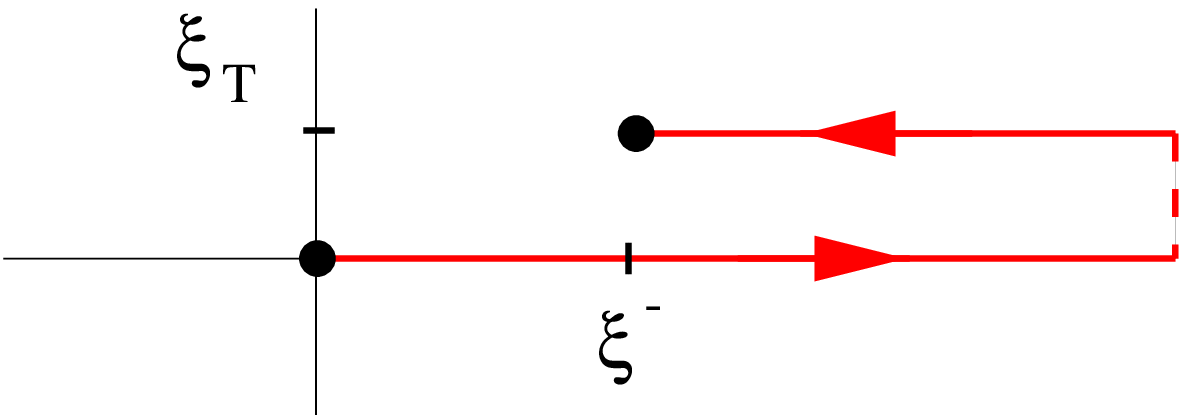}	
\end{minipage}
\end{center}
appear of the form
\be
{\cal U}(\xi,\eta) = {\cal P}\,\left. e^{-ig\int_\xi^\eta d\zeta^- A^+(\zeta)}
\right|_{\xi_\st=\eta_\st=\zeta_\st}.
\ee
The integrated results become
\bea
&&\Phi_{ij}(x) =
\int \frac{{d\xi^-}}{4\pi}\ e^{ip\cdot \xi}
\,\langle P,S\vert {\overline \psi_j(0)}
\,{\cal U}(0,\xi) \,{\psi_i(\xi)} \vert P,S\rangle \biggr|_{LC},
\\
&& \Phi_{{\partial}\,ij}^{\alpha}(x) =
\int \frac{d\xi^-}{4\pi}\ e^{ip\cdot \xi}
\,\biggl\{ \langle P,S\vert {\overline \psi_j(0)} \,{\cal U}(0,\xi)
\,{iD_T^\alpha\psi_i(\xi)} \vert P,S\rangle \biggr|_{LC}
\nonumber \\ && \quad \mbox{}
- \langle P,S\vert {\overline \psi_j(0)}\,{\cal U}(0,\infty)
\int_{\infty}^{\xi^-}d\eta^- \,{\cal U}(\infty,\eta)
\,g\,F^{+\alpha}(\eta)\,{\cal U}(\eta,\xi)\,{\psi_i(\xi)} 
\vert P,S\rangle \biggr|_{LC}\biggr\}.
\eea
We note that in $A^+ = 0$ gauge: $F^{+\alpha} = \partial^+A_T^\alpha$
from which one obtains
$A_T^\alpha(\xi) = \int_{\infty}^{\xi^-}d\eta^-\,F^{+\alpha}(\eta)$.
 
\section{Cross sections}

\vspace{1mm}
\noindent
We start with the for our purposes relevant quantities for the
DY process given in the next box. This includes the invariants
$x_A$, $x_B$ and $y$ as well as a conveniently choosen cartesian set
of vectors, which also serves to define azimuthal angles.
\begin{center}
\fbox{
\begin{minipage}{9 cm}
\epsfxsize=9.0cm \epsfbox{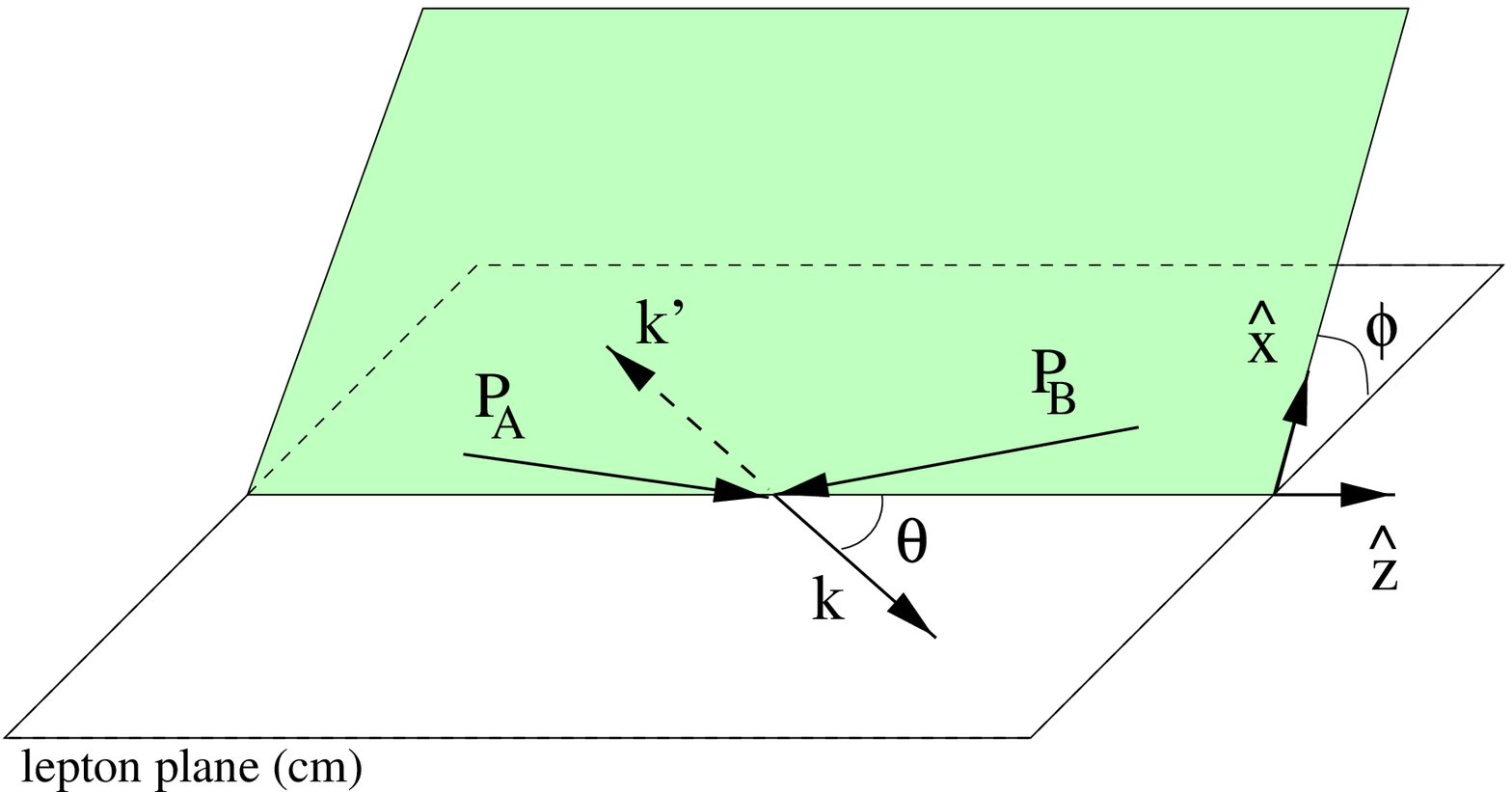}
\\
Cartesian set
\begin{eqnarray*}
&&T^\mu = q^\mu \\
&&Z^\mu = x_A P^\mu_A - x_B P^\mu_B \\
&&X^\mu = q^\mu - x_A P^\mu_A - x_B P^\mu_B 
\end{eqnarray*}
\begin{minipage}{7 cm}
\small
Note: $T^2 = Q^2$, $Z^2 = -Q^2$, $X^2 = -Q_T^2$
\end{minipage}
\end{minipage}
\begin{minipage}{6 cm}
\[
AB \longrightarrow \ell \bar\ell X 
\]
\begin{eqnarray*}
&&x_A = \frac{Q^2}{2P_A\cdot q} \\
&&x_B = \frac{Q^2}{2P_B\cdot q} \\
&&y = \frac{1}{2}\left( 1 + \cos \theta\right)\\
&& \qquad \approx \frac{k\cdot P_A}{q\cdot P_A}
\approx \frac{k^\prime\cdot P_B}{q\cdot P_B}
\end{eqnarray*}
\end{minipage}
}
\end{center}

We will be interested in particular in weighted cross sections involving
azimuthal angles and define
\bea
&&\left< {W} \right>_{\underbrace{P_A P_B}_\uparrow} 
\ \equiv\ \int d\phi^\ell\,d^2\bm q_\st\underbrace{\ {W}\ }_\uparrow
\,\frac{d\sigma_{P_A P_B}}{dx_A\,dx_B\,dy\,d\phi^\ell\,d^2\bm q_\st} ,
\\
&& \hspace{1 cm}(O,L,T)\hspace{2 cm}
{W}(Q_T,\phi_h^\ell,\phi_{S_A}^\ell,\phi_{S_B}^\ell)
\eea
where $O$, $L$, $T$ indicate unpolarized, longitudinally polarized
($\lambda$ = 1) or transversely polarized ($\vert \bm S_\st\vert$ = 1) hadrons.
The azimuthal angles $\phi_h^\ell = \phi_h - \phi^\ell$ are defined w.r.t.
the lepton plane.

One has e.g. for the simplest unpolarized $\bm q_\st$-averaged cross section
the well-known result
\be
\left< {1}\right>_{OO} 
\ =\ \frac{\pi \alpha^2}{3Q^2}
\left\lgroup 1 + \cos^2\theta \right\rgroup 
\,\sum_{a,\bar a} e_a^2
{f^a_1}(x_A) \,{f^{\bar a}_1}(x_B),
\ee
where we have included the summation over quarks and antiquarks and the
flavor indices on the distribution functions.

We start with a discussion of {\em spin asymmetries in polarized 
$\bqt$-integrated DY scattering}, where one finds the following
nonvanishing asymmetries,
\bea
&&
\left< {1} \right>_{LL}
= \frac{\pi \alpha^2}{3Q^2}\,{\lambda_A\,\lambda_B}
\left\lgroup 1 + \cos^2 \theta \right\rgroup
\sum_{a,\bar a} e_a^2
\,{g_{1}^{a}}(x_A) {g_{1}^{\bar a}}(x_B),
\\&&
\left< {\cos \left(\phi_{S_A}^\ell + \phi_{S_B}^\ell\right)} \right>_{TT}
= \frac{\pi \alpha^2}{6Q^2}\,{\lambda_A\,\lambda_B}
\left\lgroup 1 - \cos^2 \theta \right\rgroup
\sum_{a,\bar a} e_a^2
\,{h_{1}^{a}}(x_A) {h_{1}^{\bar a}}(x_B),
\\&&
\left< {\cos \left(\phi_{S_B}^\ell\right)} \right>_{LT}
= \frac{\pi \alpha^2}{3Q^2}\,{\vert \bm S_{A\,\st}\vert\,
\vert \bm S_{B\,\st}\vert}\,\sin 2\theta
\ \sum_{a,\bar a} e_a^2\,\Biggl\{
\,\frac{M_A x_A}{Q}\left({h_L^a+\tilde h_L^a}\right)
(x_A) {h_{1}^{\bar a}}(x_B)
\nonumber \\ &&\hspace{8.5 cm} \mbox{}
+ \frac{M_B x_B}{Q} \,{g_1^a}(x_A)
\left({g_T^{\bar a}+\tilde g_T^{\bar a}}\right) (x_B) \Biggr\}.
\eea
These asymmetries have been given in Ref.~\cite{JJ92} disregarding
$\bkt$-dependence of the distribution functions. In that case one has for the 
twist-three functions $\tilde g_T$ = $g_T$ and $\tilde h_L$ = $h_L$, which 
simplifies the above LT asymmetry. However, taking into account the transverse 
momentum dependence, both the interaction-dependent (tilde) functions and the 
full twist-three functions including a twist two part appear in the
LT asymmetry~\cite{TM94}. 

As a next example, we consider two specific {\em azimuthal asymmetries in 
polarized DY scattering (at leading order)}, namely
\bea
&&
\left< {\frac{Q_T}{M_B}
\,\cos \left(\phi_h^\ell - \phi_{S_B}^\ell\right)} \right>_{LT}
= \frac{\pi \alpha^2}{3Q^2}\,{\lambda_A\,\vert \bm S_{B\,\st}\vert}
\left\lgroup 1 + \cos^2 \theta \right\rgroup
\sum_{a,\bar a} e_a^2
\,{g_{1}^{a}}(x_A) {g_{1T}^{(1)\bar a}}(x_B),
\\&&
\left< {\frac{Q_T}{M_A}
\,\cos \left(\phi_h^\ell + \phi_{S_B}^\ell\right)} \right>_{LT}
= \frac{\pi \alpha^2}{3Q^2}\,{\lambda_A\,\vert \bm S_{B\,\st}\vert}
\left\lgroup 1 - \cos^2 \theta \right\rgroup
\sum_{a,\bar a} e_a^2
\,{h_{1L}^{\perp (1)a}}(x_A) {h_1^{\bar a}}(x_B),
\eea
discussed in Ref.~\cite{TM95}. Together with the previous subleading 
LT asymmetries these allow tests of the relations in Eqs~\ref{rel1}
and \ref{rel2}. We note that the constraints on the parametrization of
the quark-quark correlation functions, notably time reversal invariance
only allow double spin asymmetries in DY scattering.

\section{Time reversal symmetry}

\vspace{1mm}
\noindent
Non-applicability of time reversal symmetry (T) allows a richer structure for 
$\Phi$. This is the case in, for instance, the corresponding definitions for 
fragmentation functions~\cite{JJ93,MT95}, where one cannot use T as a symmetry 
because the final state hadron is not a plane wave state.
But also for distribution functions it has been argued that time reversal
odd quantities may appear, e.g. through soft factorization
breaking mechanisms, the chiral structure of 
nucleons or gluonic poles~\cite{S90,ABM95,BMT97}.
In the quark-quark correlation function then the following structures are
allowed
\be
\Phi(x,\bm p_\st) \biggr|_{T-odd} =
\frac{1}{2} \Biggl\{
{f_{1T}^\perp}(x,\bkt)\, \frac{\epsilon_{\mu \nu \rho \sigma}
\gamma^\mu n_+^\nu k_\st^\rho {S_\st^\sigma}}{M}
+ {h_1^\perp}(x,\bkt)\,\frac{i [\pslash_\st, \nslash_+]}{2M}
\Biggr\} .
\ee
Integrating over $\bpt$ this T-odd part vanishes, but the $\bpt$-weighted
result is nonvanishing,
\be
\Phi_{\partial}^{{\alpha}}(x)\Biggr|_{T-odd} =
\frac{M}{2}\,\Biggl\{
-{f_{1T}^{\perp (1)}}(x)
\,\epsilon^{\alpha}{}_{\mu\nu\rho}\gamma^\mu n_-^\nu {S_\st^\rho}
- {h_1^{\perp (1)}}(x)
\,\frac{i[\gamma^\alpha, \nslash_+]}{2}
\Biggr\}.
\ee
We note that, as in the T-even situation, the twist-three T-odd distribution
functions showing up in the order ($M/P^+$) part of $\Phi$,
\be
\Phi(x)\biggr|_{{T-odd}} =
\frac{M}{2P^+}\Biggl\{
{f_T}(x)\,\epsilon_\st^{\rho\sigma}{S_{T\rho}}\gamma_\sigma
-{\lambda}{e_L}(x)\,i\gamma_5 
+ {h}(x)\,\frac{i\,[\nslash_+,\nslash_-]}{2} 
\Biggr\},
\ee
can be related to the $\bkt^2$-moments in $\Phi_\partial^\alpha$,
\bea
&&{h}(x) = -\frac{d}{dx}{h_1^{\perp (1)}},
\label{rel3}
\\ &&
{f_T}(x) = -\frac{d}{dx}{f_{1T}^{\perp (1)}}.
\label{rel4}
\eea

At ${\cal O}(1/Q)$ one again needs the quark-quark-gluon correlation 
functions and the use of the equations of motion to rewrite
\be
\Phi_{{D}}^{{\alpha}}(x)\Biggr|_{T-odd} =
\frac{M}{2}\,\Biggl\{ x{f_T}(x)
\,\epsilon^{\alpha}_{\ \ \mu\nu\rho}\gamma^\mu n_-^\nu {S_\st^\rho}
+ x\,{h}(x) \,\frac{i[\gamma^\alpha, \nslash_+]}{4}
\Biggr\}.
\ee
The interaction-dependent parts can again be isolated,
\be
\Phi_{{A}}^{{\alpha}}(x)\Biggr|_{T-odd} =
\frac{M}{2}\,\Biggl\{
x{\tilde f_T}(x) 
\,\epsilon^{\alpha}_{\ \ \mu\nu\rho}\gamma^\mu n_-^\nu {S_\st^\rho}
+ x\,{\tilde h}(x) \,\frac{i[\gamma^\alpha, \nslash_+]}{4}
\Biggr\},
\ee
with
\bea
&&
x{\tilde f_T} = x{f_T} + {f_{1T}^{\perp(1)}},
\\ &&
x{\tilde h} = x{h} + 2 {h_{1}^{\perp(1)}}.
\eea
The equivalence of the Wandzura-Wilczek parts, obtained by combining
$\Phi_A^\alpha$ = 0 with Eqs~\ref{rel3} and \ref{rel4} is 
simply $f_T(x) = h(x) = 0$.

As an example of T-odd distributions we mention the following
{\em azimuthal asymmetry for unpolarized leptoproduction (leading twist)}
\be
\left<{\frac{Q_T^2}{MM_h} \,\cos(2\phi^\ell_h)}\right>_{OOO}
= \frac{16\pi \alpha^2\,s}{Q^4} \,(1-y)
\sum_{a,\bar a} e_a^2
\,\xbj\,{h_1^{\perp(1)a}}(\xbj)\,{H_1^{\perp (1)a}}(z_h),
\ee
which involves besides the T-odd distribution function $h_1^{\perp (1)}$,
depending on $\xbj$ = $Q^2/2P\cdot q$,
the equivalent fragmentation function $H_1^{\perp (1)}$ depending on $z_h$ = 
$P\cdot P_h/P\cdot q$. Examples of single spin asymmetries in the DY process
are given in the talk of Boer~\cite{BMT97}.

\section{Conclusions}

\vspace{1mm}
\noindent
I have reviewed the systematic parametrization of quark-quark and 
quark-quark-gluon correlation functions which appear as the soft parts
in the diagrammatic expansion of hard processes such as DY scattering. In
particular, I have emphasized the role of transverse momenta of quarks,
accounting for which is essential in processes like DY scattering or
1-particle inclusive lepton-hadron scattering. The effects of transverse
momenta of quarks show up, for instance, in azimuthal asymmetries.
While at leading order the relevant functions in the soft parts of 
hard processes are just the quark densities, one sees at subleading order
the QCD dynamics at work inside the hadrons.

\vspace{1mm}
\noindent
This work is part of the scientific program of the foundation for Fundamental
Research on Matter (FOM) and the Dutch Organization for Scientific Research
(NWO).

\end{document}